%
%
%
\def\today{\ifcase\month\or January\or February\or March\or April\or May\or
June\or July\or August\or September\or October\or November\or December\fi
\space\number\day, \number\year}
%
%
\newcount\notenumber

\def\note{\global\advance\notenumber by 1 \footnote{$^{\the\notenumber}$}}
%
%
\newif\ifsectionnumbering
\newcount\eqnumber
\def\cleareqnumber{\eqnumber=0}
\def\numbereq{\global\advance\eqnumber by 1
\ifsectionnumbering\eqno(\the\secnumber.\the\eqnumber)\else\eqno
(\the\eqnumber)\fi}
\def\eqalinno{{\global\advance\eqnumber by 1}
\ifsectionnumbering(\the\secnumber.\the\eqnumber)\else(\the\eqnumber)\fi}
\def\name#1{\ifsectionnumbering\xdef#1{\the\secnumber.\the\eqnumber}
\else\xdef#1{\the\eqnumber}\fi}
\def\nosectionnumbering{\sectionnumberingfalse}
\sectionnumberingtrue
%
%
\newcount\refnumber

\immediate\openout1=refs.tex
\immediate\write1{\noexpand\frenchspacing}
\immediate\write1{\parskip=0pt}
\def\ref#1#2{\global\advance\refnumber by 1%
[\the\refnumber]\xdef#1{\the\refnumber}%
\immediate\write1{\noexpand\item{[#1]}#2}}
\def\tie{\noexpand~}

%
%
\font\twelvebf=cmbx10 scaled \magstep1
\newcount\secnumber

\def\newsection#1.{\ifsectionnumbering\cleareqnumber\else\fi%
	\global\advance\secnumber by 1%
	\bigbreak\bigskip\par%
	\line{\twelvebf \the\secnumber. #1.\hfil}\nobreak\medskip\par\noindent}
%
%
%
\def \sqr#1#2{{\vcenter{\vbox{\hrule height.#2pt
	\hbox{\vrule width.#2pt height#1pt \kern#1pt
		\vrule width.#2pt}
		\hrule height.#2pt}}}}

%
%
%
\newdimen\fullhsize
\def\fiddle{\fullhsize=6.5truein \hsize=3.2truein}
\def\fullline{\hbox to\fullhsize}
\def\mkhdline{\vbox to 0pt{\vskip-22.5pt
	\fullline{\vbox to8.5pt{}\the\headline}\vss}\nointerlineskip}
\def\mkftline{\baselineskip=24pt\fullline{\the\footline}}
\let\lr=L \newbox\leftcolumn
\def\twocolumns{\fiddle
	\output={\if L\lr \global\setbox\leftcolumn=\columnbox
		\global\let\lr=R \else \doubleformat \global\let\lr=L\fi
		\ifnum\outputpenalty>-20000 \else\dosupereject\fi}}
\def\doubleformat{\shipout\vbox{\mkhdline
		\fullline{\box\leftcolumn\hfil\columnbox}
		\mkftline} \advancepageno}
\def\columnbox{\leftline{\pagebody}}
\nosectionnumbering
\magnification=1200
\def\pr#1 {Phys. Rev. {\bf D#1\tie }}
\def\pe#1 {Phys. Rev. {\bf #1\tie}}
\def\pre#1 {Phys. Rep. {\bf #1\tie}}
\def\pl#1 {Phys. Lett. {\bf #1B\tie }}
\def\prl#1 {Phys. Rev. Lett. {\bf #1\tie }}
\def\np#1 {Nucl. Phys. {\bf B#1\tie }}
\def\ap#1 {Ann. Phys. (NY) {\bf #1\tie }}
\def\cmp#1 {Commun. Math. Phys. {\bf #1\tie }}
\def\imp#1 {Int. Jour. Mod. Phys. {\bf A#1\tie }}
\def\mpl#1 {Mod. Phys. Lett. {\bf A#1\tie}}
\def\tie{\noexpand~}

\parskip=15pt plus 4pt minus 3pt
\headline{\ifnum \pageno>1\it\hfil Supersymmetric Sum
Rules	$\ldots$\else \hfil\fi}
\font\title=cmbx10 scaled\magstep1
\font\tit=cmti10 scaled\magstep1
\footline{\ifnum \pageno>1 \hfil \folio \hfil \else
\hfil\fi}
\raggedbottom


\def\half{{\textstyle{1\over2}}}
\def\ha{{1\over2}}
\overfullrule0pt


\rightline{\vbox{\hbox{RU98-04-B}\hbox{NYU-TH/98/03/01}\hbox{hep-th/9803073}}}
\vfill
\centerline{\title SUPERSYMMETRIC SUM RULES FOR ELECTROMAGNETIC}
\centerline{\title MULTIPOLES}
\vfill
{\centerline{\title Ioannis Giannakis${}^{a}$,
James T.~Liu${}^{a}$ and Massimo Porrati${}^{a, b}$ \footnote{$^{\dag}$}
{\rm e-mail: \vtop{\baselineskip12pt
\hbox{giannak@theory.rockefeller.edu, jtliu@theory.rockefeller.edu,}
\hbox{massimo.porrati@nyu.edu}}}}
}
\medskip
\centerline{$^{(a)}${\tit Physics Department, The Rockefeller
University}}
\centerline{\tit 1230 York Avenue, New York, NY
10021-6399}
\medskip
\centerline{$^{(b)}${\tit Department of Physics, New York University}}
\centerline{\tit 4 Washington Pl., New York, NY 10003} 
\vfill
\centerline{\title Abstract}
\bigskip
{\narrower\narrower
We derive model independent, non-perturbative supersymmetric
sum rules for the magnetic and electric 
multipole moments of any theory with $N=1$ supersymmetry.
We find that in any irreducible $N=1$ supermultiplet the diagonal matrix 
elements of the $l$-multipole moments are completely fixed in terms of their
off-diagonal matrix elements and the diagonal $(l-1)$-multipole moments.
\par}
\vfill\vfill\break


\newsection Introduction.%
Supersymmetry imposes constraints on the magnetic moments
of the particle states
\ref\por {S. Ferrara and M. Porrati, Phys. Lett. {\bf B288} (1992) 85.},
\ref\gia{I. Giannakis and J. T. Liu, Rockefeller University preprint,
RU-97-08-B,\goodbreak {\tt hep-th/9711173}.}.
These constraints are model independent,
valid for any massive $N=1$ and $N=2$ supermultiplet. They are also
in agreement with the results of Ferrara and Remiddi
\ref\ferrem{S. Ferrara and E. Remiddi, Phys. Lett. {\bf 53B} (1974) 347.},
who showed that $g=2$ to all orders in perturbation theory for any
$N=1$ chiral multiplet and those of Bilchak, Gastmans and Van~Proeyen
\ref\bgvan{C. L. Bilchak, R. Gastmans and A. Van Proeyen, Nucl. Phys.
{\bf B273} (1986) 46.}
who demonstrated that when spin-$1$ fields
are present supersymmetry does not necessarily demand $g=2$, but
nevertheless leads to a relation between the $g$-factors of the spin-1/2
and spin-1 particles of the superspin~1/2 multiplet.

The model independent magnetic dipole moment sum rules were derived
in [\por] by noting that supersymmetry relates the matrix elements of the
conserved electromagnetic current within the various states of a general
massive supermultiplet.  By selecting the magnetic dipole term in the
multipole expansion of the electromagnetic current, the authors of [\por]
found, for the gyromagnetic ratios, the following sum rule:
$$
\eqalign{g_{j+\ha} &= 2 + 2 j h_j,\cr
g_j &= 2 + (2j+1) h_j,\cr
g_{j-\ha} &= 2 + (2j+2) h_j.}
\numbereq\name{\gfactors}
$$
Note that $j$ is the superspin labeling the massive supermultiplet which
contains states of spins $(j+\ha, j, j, j-\ha)$.  Since both spin-$j$ states
have identical gyromagnetic ratios, we see that all the $g$-factors are
determined in terms of a single real number, $h_j$, corresponding to an
off-diagonal magnetic dipole matrix element between the $j+\ha$ and $j-\ha$
states of the supermultiplet. In the special cases $j=0,1/2$, the sum rules
read
$$ 
\matrix{g_\ha = 2& (j=0),\cr 
g_1 = 2 + h_\ha ,\;\; g_\ha=2 +2h_\ha &(j=\half).}
\numbereq\name{\specialcase}
$$
Notice that chiral multiplets ($j=0$) have a fixed gyromagnetic ratio $g=2$.

In this letter, we generalize the above gyromagnetic ratio sum rule to
encompass higher multipole moments (both electric and magnetic).  This is
easily done by working to all orders in the momentum transfer in the
appropriate electromagnetic matrix elements.  The resulting
multipole sum rules have a similar structure as (\gfactors), and take
the form
$$
\eqalign{
{\cal T}_{j+{1\over 2}}^{(l)^{(e, m)}}&={\mp}{1\over M}
{\cal T}_{j}^{(l-1)^{(m, e)}}+{{2j+1-l}\over l}
{\cal H}_{j}^{(l)^{(e, m)}},\cr
{\cal T}_{j}^{(l)^{(e, m)}}&={\mp}{1\over M}
{\cal T}_{j}^{(l-1)^{(m, e)}}+{{2j+1}\over l}
{\cal H}_{j}^{(l)^{(e, m)}},\cr
{\cal T}_{j-{1\over 2}}^{(l)^{(e, m)}}&={\mp}{1\over M}
{\cal T}_{j}^{(l-1)^{(m, e)}}+{{2j+1+l}\over l}
{\cal H}_{j}^{(l)^{(e, m)}},\cr}
\numbereq\name{\sumrule}
$$
%
\def\defT{21}%
where the electric/magnetic $l$-pole generalization of $g$ is
denoted by ${\cal T}^{(l)^{(e,m)}}_j$ and is defined in Eqn.~(\defT) below%
\note{ More precisely these sum rules hold for the generic case
$2j\ge {l+1}$, where $j$ denotes the superspin. Note that
${\cal T}^{(l)^{(e,m)}}_j$ is meaningless whenever $l>2j$, as may be
infered from Eqn.~(\defT).}.
These sum rules indicate the general
structure imposed by supersymmetry that the electric (magnetic) $l$-pole
moments are completely determined solely in terms of a single magnetic
(electric) $(l-1)$-pole moment and the real quantity ${\cal H}^{(l)}$
parameterizing an off-diagonal transition between the spin $j\pm\ha$ states
of the multiplet.
Note that the upper and lower signs in (\sumrule) and subsequent
equations correspond to the first and second entries in {\it e.g.} $(e,m)$.
This difference in sign between the electric and magnetic sum rules may be
understood intuitively from electromagnetic duality which exchanges
electric and magnetic fields, ${\vec E}\to {\vec B}$
and ${\vec B}\to -{\vec E}$.

When $l=1$, the magnetic part of the sum rule (\sumrule) reduces to the
result of Ferrara and Porrati, (\gfactors), since $g$ is defined as a ratio:
${\cal T}_j^{(1)^{(m)}} = {g_j e\over 2M}$ and ${\cal T}_j^{(0)^{(e)}} = e$.
Furthermore, just as for the magnetic dipole moments, we note that setting
${\cal H}_j^{(l)^{(e,m)}} = 0$ yields the ``preferred'' value for the
$l$-pole moments
$$
{\cal T}^{(l)^{(e,m)}}_{j+\ha} = {\cal T}^{(l)^{(e,m)}}_j =
{\cal T}^{(l)^{(e,m)}}_{j-\ha} =
\mp {1\over M} {\cal T}^{(l-1)^{(m,e)}}_j,
\numbereq
$$
generalizing the notion of $g=2$ as the preferred value of the gyromagnetic
ratio.  

\newsection Derivation of the Sum Rules.%
Derivation of the sum rules (\sumrule) follows the method of [\por], and
involves the transformation properties of a conserved current $J_\mu$ that
commutes with the $N=1$ supersymmetry algebra.  The main complication in
obtaining the present results is the requirement of working to all orders
in the multipole expansion (and as a result having to keep track of
higher-order in momentum transfer terms in the matrix elements).

Recall that the $N=1$ algebra has the form
$\lbrace Q_{\alpha}, {\overline Q}_{\beta} \rbrace
= 2({\gamma}^{\mu})_{\alpha\beta}
P_{\mu}$, where ${\overline Q}=Q^{T}C$ is the Majorana conjugate
and $C$ is the charge conjugation matrix obeying
$C{\gamma^{\mu}}C^{-1}=-{\gamma^{\mu T}}$ and $C^2=-1$.  For a massive
single particle state, we may work in the rest frame $P^{\mu}=(M,0,0,0)$.
Defining chiralities
$$
{\gamma_5}Q^{L\atop R}=\pm Q^{L\atop R},
\numbereq
$$
and helicities
$$
{\gamma^{12}}Q_{\pm{1\over 2}}={\mp i}Q_{\pm {1\over 2}},
\numbereq
$$
the supersymmetry algebra can be recast as follows:
$$
\lbrace Q^{L}_{1\over 2}, Q^{R}_{-{1\over 2}} \rbrace=2M,
\qquad
\lbrace Q^{L}_{-{1\over 2}}, Q^{R}_{1\over 2} \rbrace=2M,
\numbereq\name{\eqana}
$$
while the remaining anticommutators vanish%
\note{To fix our phase conventions, we work in the Dirac
representation for the ${\gamma}$-matrices and take
$\gamma_5=i{\gamma^{0}}{\gamma^{1}} {\gamma^{2}}{\gamma^{3}}$ and
$C=i{\gamma^{0}}{\gamma^{2}}$.  The spinors then decompose as
${\sqrt 2}Q_{\alpha}^T=Q^{L}_{{1\over 2}}[1\ 0\ 1\ 0]
+Q^{L}_{-{1\over 2}}[0\ 1\ 0\ 1]
+Q^{R}_{{1\over 2}}[-1\ 0\ 1\ 0]
+Q^{R}_{-{1\over 2}}[0\ 1\ 0\ -1]$.}.
We may rescale the supercharges according to
$q^{L,R}_{{\pm}{1\over 2}}={1\over\sqrt{2M}}
Q^{L, R}_{{\pm}{1\over 2}}$ to
recover the Clifford algebra for two fermionic degrees of freedom.
{} One can then construct
its irreducible representations by starting with a superspin $j$
Clifford vacuum, $|j\rangle$, annihilated by $q^{L}_{{\pm}{1\over 2}}$,
and acting on it with the creation operators $q^{R}_{{\pm}{1\over 2}}$.
As a result, we see that the representation has dimension
$(2j+1)\times2^2$ where $2j+1$ is the degeneracy of the original spin $j$
state..  The spins of the states are given by the addition of angular
momenta, $j \times [(1/2) + 2(0)]$, giving
states of spins $j-{1\over 2}$, $j$ and  $j+{1\over 2}$
with degeneracies $1, 2, 1$.

Since the supercharges $Q^{L, R}_{\pm{1\over 2}}$
are operators of spin $1/2$, this leads to a
shorthand notation for labeling the states of a massive $N=1$
multiplet in the following manner:
the spin $j$ Clifford vacuum is denoted by $|0\rangle$,
acting
on this state with the normalized supercharges $q^{R}_{{1\over 2}}$ or
$q^{R}_{-{1\over 2}}$ then results in the spin `up' or `down' states
$|{\uparrow}\rangle$ or $|{\downarrow}\rangle$ respectively.
The action of two $q$'s on the Clifford vacuum is
denoted by $|{\updownarrow}\rangle$.

For $N=1$ supersymmetry, any conserved current commuting
with the supersymmetry generators must belong to a real linear
multiplet. The components of a real linear multiplet multiplet
are $(C(x), \zeta(x), J_{\mu}(x))$, where $C(x)$ is a real scalar
and $\zeta(x)$ a Majorana spinor. As a result of current
conservation, ${\partial^{\mu}}J_{\mu}=0$, the multiplet consists
of $4$ fermionic and $4$ bosonic degrees of freedom.
The transformation properties of the
components under a supersymmetry variation are given by
$$
{\delta}C=i {\overline{\epsilon}}{\gamma_5}{\zeta}, \quad
{\delta}{\zeta} =i({\gamma^{\lambda}}J_{\lambda}+i{\gamma_5}{\gamma^{\lambda}}
{\partial_{\lambda}}C){\epsilon}, \quad
{\delta}J_\mu=-\overline{\epsilon}\gamma_\mu{}^\lambda\partial_\lambda\zeta.
\numbereq\name{\eqgian}
$$
It follows that two successive supersymmetry transformations on the
conserved current $J_{\mu}$ gives
$$
{\delta}_{\eta}{\delta}_{\epsilon}J_{\mu}=
i{\overline{\epsilon}}\gamma_\mu{}^\nu\gamma^\rho
(\partial_\nu J_\rho-i\gamma_5 \partial_\nu\partial_\rho C)\eta.
\numbereq\name{\eqpasa}
$$
The matrix elements of this equation between single particle states
which belong to the
same $N=1$ multiplet give rise to sum rules for the electromagnetic
multipoles of the particle states.

To obtain the connection between the matrix elements of $J_\mu$ and the
terms in the multipole expansion, we first recall the standard definitions
(see {\it e.g.}
\ref\jackson{J. D. Jackson, {\it Classical Electrodynamics}, pp. 755--758,
Wiley, New York (1975).})
for the electric $l$-pole moments
$$
Q^{(l)}_{i_1i_2\cdots i_l} = \int d^3x (x_{i_1}x_{i_2}\cdots x_{i_l}) J_0(x)
- \hbox{trace},
\numbereq
$$
and the magnetic $l$-pole moments
$$
M^{(l)}_{i_1i_2\cdots i_l} = -{1\over l+1}\int d^3x
(x_{i_1}x_{i_2}\cdots x_{i_l}) \vec\nabla \cdot (\vec x \times \vec J(x))
-\hbox{trace}.
\numbereq
$$
While ordinarily defined in terms of spherical tensors
(see {\it e.g.}
\ref\ren{V. Rahal and H. C. Ren, \pr41 (1989) 1989.}),
the above multipole
moments, expressed as cartesian tensors, are more naturally related to the
expansions for the matrix elements of $J_{\mu}$,
$$
\eqalign{
\langle j', m', \vec p\,|J_{0}|j, m, 0\rangle&=
\sum_{l=0}^{\infty}{1\over {l!}}(ip)_{i_1}(ip)_{i_2} \cdots
(ip)_{i_l}\langle j', m', 0,|T_{{i_1}{i_2} \cdots {i_l}}^{(l)^{e}}
|j, m, 0\rangle,\cr
\langle j', m', \vec p\,|J_{i}|j, m, 0\rangle&=
p_{i}\langle j', m', 0,|{\Lambda}|j, m, 0\rangle\cr
&-i{\epsilon_{ijk}}p_{j}\sum_{l=1}^{\infty}{1\over l!}
(ip)_{i_2}(ip)_{i_3} \cdots
(ip)_{i_l}\langle j', m', 0,|T_{k{i_2} \cdots {i_l}}^{(l)^{m}}
|j, m, 0\rangle. \cr}
\numbereq\name{\eqnini}
$$
In particular, the {\it traceless} components of $T^{(l)^{(e)}}$ and
$T^{(l)^{(m)}}$ correspond exactly to $Q^{(l)}$ and $M^{(l)}$ respectively.
Note that the matrix elements of $\Lambda$ are completely determined by
current conservation, $\Lambda = -\left({E-M\over p^2}\right) J_0$.

The multipole moment sum rules are derived by taking the
double supersymmetry variation of the conserved current $J_\mu$,
$$
\eqalign{
\delta_\eta\delta_\epsilon J_\mu &=
[\overline{\eta} Q, [\overline{\epsilon}Q, J_\mu ]]\cr
&=\overline{\eta}Q\overline{\epsilon}QJ_\mu
-\overline{\eta}QJ_\mu \overline{\epsilon}Q
-\overline{\epsilon}QJ_\mu \overline{\eta}Q
+J_\mu \overline{\epsilon}Q\overline{\eta}Q,\cr}
\numbereq\name{\eqsamba}
$$
and evaluating it between single particle states $\langle\alpha|$ and
$|\beta\rangle$.  Since the supercharge $Q$ generates superpartners
($Q|\alpha\rangle \sim |\tilde\alpha\rangle$), this expression relates
matrix elements of $J_\mu$ between different states of a supermultiplet
in terms of $\delta_\eta\delta_\epsilon J_\mu$, which is given by (\eqpasa).
The electromagnetic $l$-pole sum rules then follow by using (\eqnini) to
expand the matrix elements in terms of multipoles and then by collecting
terms of order $p^l$.  We note that an important simplification occurs since
we are only interested in sum rules on the static multipole moments.  This
means in practice that all terms depending explicitly on the contracted
momentum $p^2$ may be ignored, as they do not contribute to the static
$l$-pole moments (and instead correspond to the trace terms in
$T^{(l)^{(e,m)}}$)%
\note{In principle supersymmetry would give complete relations between
electromagnetic form factors $T^{(l)^{(e,m)}}(p^2)$ of superpartners.
However in this case it appears the moments of the ``auxiliary field'' $C$
enter in a non-trivial manner.}.

The general double supersymmetry variation procedure is simplified in
practice by choosing the global supersymmetry transformation parameters
$\eta$ and $\epsilon$ in such a way that several terms on the right hand
side of (\eqsamba) act as annihilation operators on the initial or final
states and hence may be dropped. In particular, by choosing $\eta_L=0$,
we find
$$
\langle\alpha, \vec p\,|\delta_{\eta_R}\delta_\epsilon J_\mu|\beta, 0\rangle
= \langle\alpha,\vec p\,|J_\mu\overline{\epsilon}Q\overline{\eta}_RQ
|\beta,0\rangle
- \langle\alpha,0|\overline{\epsilon}Q^{(p)}L^{-1}(\vec p\,)J_\mu
\overline{\eta}_RQ|\beta,0\rangle,
\numbereq\name{\gensum}
$$
where $Q^{(p)}$ denotes the Lorentz boost of $Q$, namely
$Q^{(p)}=L^{-1}(\vec p\,)QL(\vec p\,)$,
and $|\alpha, \vec p\ \rangle=L(\vec p\,)|\alpha, 0\rangle$.

By further choosing $\epsilon_L=0$, and noting from (\eqpasa) that
$\delta_{\eta_R}\delta_{\epsilon_R}J_\mu=0$, we easily obtain the
``vanishing'' sum rule,
$$
\langle\alpha,\vec p\,|J_\mu\overline{\epsilon}_RQ\overline{\eta}_RQ
|\beta,0\rangle = 0.
\numbereq
$$
This demonstrates that {\it all} matrix elements of the electromagnetic
current vanish between states $|0\rangle$ and
$\left|\updownarrow\right\rangle$, and
hence that there are no off-diagonal moments between the two spin-$j$
states of the supermultiplet.

If instead we choose $\epsilon_R=0$ and make use of the fact that $Q$
transforms as a spinor,
$$
Q^{(p)}=e^{{1\over 2}{\omega^i}{\gamma^{0i}}}
Q={\sqrt{{E+M}\over {2M}}}(I+{p^i\over E+M} \gamma^{0i})Q,
\numbereq\name{\eqnvcus}
$$
we obtain from (\gensum) the expression
$$
\eqalign{
\langle\alpha,\vec p\,|i\overline{\epsilon}_{L}{\gamma}_\mu{}^\nu
{\gamma^\lambda}{\partial_\nu}(J_{\lambda}-i{\gamma_5}{\partial_\lambda}C)
{\eta_{R}}&|\beta,0 \rangle\cr
&=2M(\overline{\epsilon}_{L}{\gamma^0}{\eta_{R}})
\langle\alpha,\vec p\,|J_\mu|\beta,0\rangle\cr
&\quad-{\sqrt{{E+M}\over {2M}}}\langle\alpha,0|
\overline{\epsilon}_{L}QL^{-1}(\vec p\,)J_\mu\overline{\eta}_{R}Q
|\beta,0\rangle\cr
&\quad-{{p^i}\over {\sqrt{2M(E+M)}}}\langle\alpha,0|
\overline{\epsilon}_{L}\gamma^{0i}QL^{-1}(\vec p\,)J_{\mu}
\overline{\eta}_{R}Q|\beta,0\rangle.\cr}
\numbereq\name{\eqasxuti}
$$
Equation (\eqasxuti) can be simplified significantly if we ignore $p^2$
({\it i.e.}~trace) terms which do not contribute to the electromagnetic
multipole sum rules.  After some manipulation, the time and space
components of Eqn.~(\eqasxuti) can be written as follows:
$$
\eqalign{
{1\over {2M}}\langle\alpha,0|
\overline{\epsilon}_{L}QL^{-1}(\vec p\,)J_0\overline{\eta}_{R}Q
|\beta,0\rangle&=(\overline{\epsilon}_{L}{\gamma^0}{\eta_{R}})
\langle\alpha,\vec p\,|J_0|\beta,0\rangle-i{\epsilon_{ijk}}
{{p^j}\over {2M}}(\overline{\epsilon}_{L}{\gamma^k}
{\eta_{R}})\langle\alpha, \vec p\,|J_i|\beta,0\rangle,\cr
{1\over {2M}}\langle\alpha,0|
\overline{\epsilon}_{L}QL^{-1}(\vec p\,)J_i\overline{\eta}_{R}Q
|\beta,0\rangle&=(\overline{\epsilon}_{L}{\gamma^0}{\eta_{R}})
\langle\alpha,\vec p\,|J_i|\beta,0\rangle-i{\epsilon_{ijk}}
{{p^j}\over {2M}}(\overline{\epsilon}_{L}{\gamma^k}
{\eta_{R}})\langle\alpha, \vec p\,|J_0|\beta,0\rangle,\cr}
\kern-12pt
\numbereq\name{\eqcuidl}
$$
where we have omitted terms explicitly proportional to $p^2$.
Note in particular that matrix elements of $C$ do not enter.

We now use the explicit multipole expansion of the matrix elements,
(\eqnini), and equate terms of the same order in $\vec p$.  Because of the
explicit factor of $p^j$ in (\eqcuidl), we see that multipole terms of order
$l$ and $l-1$ are explicitly related; heuristically Eqn.~(\eqcuidl) states
that the $l$-pole moment of a superpartner is given by the same $l$-pole
moment of the original state plus a correction based on the opposite
(electric/magnetic) $(l-1)$-pole.  Explicitly, at order
$p_{i_1}p_{i_2} \cdots p_{i_l}$, we find
$$
\eqalign{
{1\over {2M}}\langle\alpha,0|
\overline{\epsilon}_{L}QT_{{i_1}{i_2} \cdots {i_l}}^{(l)^{(e, m)}}
\overline{\eta}_{R}Q
|\beta,0\rangle&=
(\overline{\epsilon}_{L}{\gamma^0}{\eta_{R}})
\langle\alpha,0|T_{{i_1}{i_2} \cdots {i_l}}^{(l)^{(e, m)}}
|\beta,0\rangle\cr
&\quad{\mp}{l\over {2M}}
(\overline{\epsilon}_{L}{\gamma^{i_1}}
{\eta_{R}})\langle\alpha, 0|T_{{i_2}{i_3} \cdots {i_l}}^{(l-1)^{(m, e)}}
|\beta,0\rangle\cr
&\quad{\pm}{l\over {2M}}{{l-1}\over {2l-1}}{\delta_{i_{1}i_{2}}}
(\overline{\epsilon}_{L}{\gamma^{j}}
{\eta_{R}})\langle\alpha, 0|T_{j{i_3} \cdots {i_l}}^{(l-1)^{(m, e)}}
|\beta,0\rangle,\cr}
\numbereq\name{\eqcdvdl}
$$
where the indices $i_1, i_2,\ldots, i_l$ are to be explicitly symmetrized,
and all tensor quantities are assumed traceless.
Note that the last term in (\eqcdvdl) is responsible for subtracting out the
trace from the spin-1 $\times$ spin-($l-1$) combination
$(\overline{\epsilon}_L\vec\gamma\eta_R) \times T^{(l-1)^{(m,e)}}$.

Because of rotational invariance, each $l$-pole moment may be completely
characterized by a single quantity---essentially a reduced matrix element
according to the Wigner-Eckart theorem.  In particular, for a single
particle state of spin $j$ and $z$-component $m$, we define the reduced
$l$-pole moment ${\cal T}_j^{(l)^{(e,m)}}$ by
$$
\langle j, m'| T_{i_1i_2\cdots i_l}^{(l)^{(e,m)}} | j, m \rangle
= {\cal T}_j^{(l)^{(e,m)}} \langle j, m'| ( J_{(i_1} J_{i_2}\cdots J_{i_l)}
-\hbox{trace})|j, m\rangle.
\numbereq\name{\wigner}
$$
The sum rules may now be established by
examining the $i_1,i_2, \ldots i_l = 3,3,\ldots,3$ components of (\eqcdvdl).
Furthermore, the spin-$j$ angular momenta manipulations are simplified by
picking the particular $m=j$ state in the matrix elements, in which case
(\wigner) may be reexpressed as
$$
\langle j, j| T_{33\cdots3}^{(l)^{(e,m)}} |j, j\rangle
= {[(2j)(2j-1)\cdots(2j-(l-1))] \over {2l \choose l}}
{\cal T}_j^{(l)^{(e,m)}}.
\numbereq\name{\defT}
$$
With the same motivation we define the multipole transition moments
${\cal H}_j^{(l)^{(e,m)}}$ as
$$
\langle j-\half, j-\half| T_{33\cdots3}^{(l)^{(e,m)}} | j+\half,j-\half\rangle
= {l\over\sqrt{2j}}
{[(2j)(2j-1)\cdots(2j-(l-1))] \over {2l \choose l-1}}
{\cal H}_j^{(l)^{(e,m)}}.
\numbereq
$$

Recall that in (\eqcdvdl) both $\langle \alpha, 0|$ and $|\beta, 0 \rangle$
denote the spin-$j$ Clifford vacuum state, $|j,m,0\rangle$, which may be
abbreviated as $|0\rangle$.  By choosing the spinor parameters $\eta_R$ and 
$\epsilon_L$ appropriately, we then relate the electromagnetic multipoles of
the different members of the $N=1$ massive multiplet.  With a total of 
two $\eta_R$ and two $\epsilon_L$ parameters, we find
$$
\eqalign{
\left\langle \uparrow \right|T_{33 \cdots 3}^{(l)^{(e, m)}}
\left|\uparrow \right\rangle&=
\langle 0|T_{33 \cdots 3}^{(l)^{(e, m)}}|0 \rangle
{\mp}{l\over {2M}}{{l}\over {2l-1}}\langle 0
|T_{33 \cdots 3}^{(l-1)^{(m, e)}}|0 \rangle,\cr
\left\langle \downarrow \right|T_{33 \cdots 3}^{(l)^{(e, m)}}
\left|\downarrow \right\rangle&=
\langle 0|T_{33 \cdots 3}^{(l)^{(e, m)}}|0 \rangle
{\pm}{l\over {2M}}{{l}\over {2l-1}}\langle 0
|T_{33 \cdots 3}^{(l-1)^{(m, e)}}|0 \rangle,\cr
\left\langle \uparrow \right|T_{33 \cdots 3}^{(l)^{(e, m)}}
\left|\downarrow \right\rangle&=
\phantom{
\langle 0|T_{33 \cdots 3}^{(l)^{(e, m)}}|0 \rangle
}
{\mp}{l\over {2M}}{{l-1}\over {2l-1}}\langle 0
|T_{-3 \cdots 3}^{(l-1)^{(m, e)}}|0 \rangle,\cr
\left\langle \downarrow \right|T_{33 \cdots 3}^{(l)^{(e, m)}}
\left|\uparrow \right\rangle&=
\phantom{
\langle 0|T_{33 \cdots 3}^{(l)^{(e, m)}}|0 \rangle
}
{\mp}{l\over {2M}}{{l-1}\over {2l-1}}\langle 0
|T_{+3 \cdots 3}^{(l-1)^{(m, e)}}|0 \rangle,\cr}
\numbereq\name{\eqaziosb}
$$
where $\pm$ in the indices denote the combinations $x^1\pm i x^2$, and the
states $\left|\uparrow\right\rangle$ and $\left|\downarrow\right\rangle$
are implicitly understood
in terms of the Clebsch-Gordon combination of spin-$1/2$ $\times$ spin-$j$.
This is the main result of our paper. The matrix elements of the
$l$-electric (magnetic) multipole moment between different members of
the supermultiplet are given in terms of the matrix elements
of the $l$-electric (magnetic) multipole moment and the
$(l-1)$-magnetic (electric) multipole moment between the Clifford
vacuum.

Finally by carrying out the addition of the superspin $j$ to
the supersymmetry generated spin we find the following sum rules:
$$
{\cal T}_{j+{1\over 2}}^{(l)^{(e, m)}}={\cal T}_{j}^{(l)^{(e, m)}}
-{\cal H}_{j}^{(l)^{(e, m)}}, \qquad
{\cal T}_{j-{1\over 2}}^{(l)^{(e, m)}}={\cal T}_{j}^{(l)^{(e, m)}}
+{\cal H}_{j}^{(l)^{(e, m)}},
$$
$$
{\cal H}_{j}^{(l)^{(e, m)}}={l\over {2j}}
\lbrack {\cal T}_{j}^{(l)^{(e, m)}}{\pm}{1\over M}
{\cal T}_{j}^{(l-1)^{(m, e)}} \rbrack,
\numbereq\name{\eqcoinv}
$$
which may be written in a completely equivalent form as presented in
Eqn.~(\sumrule).  Note that both spin-$j$ states carry identical $l$-pole
moments, as may be established using the same argument as in [\por].

\newsection Discussion.%
While the sum rules were derived for generic superspin $j$, it is important
to realize that angular momentum selection rules forbid both diagonal
(${\cal T}_j^{(l)}$) and non-diagonal (${\cal H}_j^{(l)}$) $l$-pole
electromagnetic moments whenever $l>2j$.  For $l=1$ (dipole moment), the
magnetic sum rule reduces to that of Ref.~[\por], while the electric sum rule
gives rise to the relation between EDM's:
$$
d_{j+\ha} = d_j - {d_j \over 2j+1},\qquad
d_{j-\ha} = d_j + {d_j \over 2j+1},
\numbereq
$$
where $\langle j, m| d^e_3 | j, m\rangle = d_j m$.

The special cases $j=0$ and $j=1/2$ are noteworthy. For $j=0$ only dipole
moments are allowed (for the spin-$1/2$ particle), in which case
the gyromagnetic ratio of the spin-1/2 particle in the supermultiplet
is $g=2$, as shown by Ferrara and Remiddi~[\ferrem]. 
For $j=1/2$ (massive vector multiplet) only dipole and quadrupole moments
are allowed.  Robinett
\ref\robinett{R. W. Robinett, Phys. Rev. {\bf D31} (1985) 1657.}
and Bilchak, Gastmans and Van Proeyen~[\bgvan] showed that the electric
quadrupole of the spin-1 particle is completely determined in terms of
its anomalous magnetic dipole moment.
Our sum rule reproduces this result. Indeed, by setting $j=1/2$ and $l=2$ in 
Eqn.~(\sumrule) we find the following relation between electric quadrupole 
and magnetic dipole:
$$
{\cal T}_1^{(2)^{(e)}} =-{1\over M}{\cal T}_{1/2}^{(1)^{(m)}}.
\numbereq\name{\eqquadr}
$$
Since the conventional quantum definition of the electric quadrupole
moment is given by $Q_j = \langle j, j| \int d^3 x (3 z^2 - r^2) J_0(x)
|j, j \rangle$, and is related to ${\cal T}^{(2)^{(e)}}$ by $Q_j = j (2j - 1)
{\cal T}_j^{(2)^{(e)}}$, the above relation may in fact be rewritten as
(cfr.~[\bgvan]):
$$
Q_1 = - (g_1 - 1) {e\over M^2}
\numbereq\name{\eqarcnd}
$$
(where the $g$-factor sum rule (\gfactors) was also used).
This result can be understood in the following way.
The action of a massive, charged vector multiplet $W$ coupled to a real, 
massless vector multiplet $V$ can be written in superfields as~[\por]:
$$
S=\left(\int d^2\theta W^+_\alpha W^{\alpha\,-} +
a\int d^4\theta WD_\alpha W^\dagger e^{-V}V^\alpha +{\rm c.c.}\right)
+M^2\int d^4\theta e^{-V}W^\dagger W.
\numbereq\name{\superf}
$$
Here $M$ is the mass of $W$ and $a$ is an arbitrary constant; $W^\pm_\alpha$
and $V_\alpha$ are defined as in~[\por].   
The term proportional to $a$ is the only superfield expression that
contributes to the magnetic dipole. Expanding in components, indeed, one finds
a term proportional to
$$
\int d^4x W^{\mu\,*}W^\nu F_{\mu\nu}.   
\numbereq\name{\contr}
$$
The magnetic-dipole contribution comes by setting $\mu,\nu=i,j$ ($i,j=1,2,3$).
On the other hand, by setting $\mu=0$, $\nu=i$ ($i=1,2,3$), one finds a
contribution to the electric quadrupole, since on shell and at low momenta
$\partial_\mu W^\mu=0 \Rightarrow MW^0\approx i\partial_iW^i$:
$$  
\int d^4x W^{0\,*}W^i F_{0i}= {i\over M}\int d^4x W^{j\,*}W^i 
\partial_j F_{0i}+\ldots \;\;\; ({\rm on\; shell)}.
\numbereq\name{\onshell}
$$
No other quadrupole term can be written in superfields; therefore, the 
electric quadrupole is completely determined by the magnetic dipole, as 
explicitly found in~[\bgvan] and implied by our sum rules.
\vskip .1in
\noindent
{\bf Acknowledgments.} \vskip .01in \noindent
We would like to thank V. P. Nair and H. C. Ren for useful discussions.
This work was supported in part by the Department of Energy under Contract
Number DE-FG02-91ER40651-TASK B, and by NSF under grant PHY-9722083.

\immediate\closeout1
\bigbreak\bigskip

\line{\twelvebf References. \hfil}
\nobreak\medskip\vskip\parskip

\input refs

\vfil\end

\bye